\documentclass[aps,prl,superscriptaddress,reprint]{revtex4-1} 

\usepackage[colorlinks=true,linkcolor=blue,citecolor=blue,urlcolor=blue]{hyperref}

\usepackage{setspace} 
\usepackage{graphicx}
\usepackage{amsmath}
\usepackage{color}
\usepackage{amsmath}
\usepackage{amssymb}
\usepackage{verbatim}
\usepackage{latexsym}
\usepackage{enumerate} 
\usepackage{bm} 
\usepackage{ulem} 
\usepackage[caption=false]{subfig}
\usepackage{multirow}
\usepackage{booktabs}
\usepackage{lineno}

\begin{document}

\title{%
Stability and strength of monolayer polymeric C$_{60}$
}

\author{Bo Peng}
\email{bp432@cam.ac.uk}
\affiliation{Theory of Condensed Matter Group, Cavendish Laboratory, University of Cambridge, J.\,J.\,Thomson Avenue, Cambridge CB3 0HE, United Kingdom}

\date{\today}

\begin{abstract}
Two-dimensional fullerene networks have been synthesized in several forms [Hou \textit{et al., Nature} \textbf{2022}, 606, 507], and it is unknown which monolayer form is stable at ambient condition. Using first principles calculations, I show that the believed stability of the quasi-tetragonal phases is challenged by mechanical, dynamic or thermodynamic stability. For all temperatures, the quasi-hexagonal phase is thermodynamically least stable. However, the relatively high dynamic and mechanical stabilities suggest that the quasi-hexagonal phase is intrinsically stronger than the other phases under various strains. The origin of the high stability and strength of the quasi-hexagonal phase can be attributed to the strong covalent C$-$C bonds that strongly hold the linked C$_{60}$ clusters together, enabling the closely packed hexagonal network. These results rationalize the experimental observations that so far only the quasi-hexagonal phase has been exfoliated experimentally as monolayers.
\end{abstract}

\flushbottom

\maketitle

Recent attempts to synthesize layers of connected buckyballs, i.e. C$_{60}$ molecules linked by carbon$-$carbon bonds, have obtained different arrangements of cluster cages through the formation of bonds between neighboring C$_{60}$ molecules\,\cite{Hou2022}. The obtained allotropes include a few-layer rectangular structure in which each C$_{60}$ molecule has four neighboring buckyballs and a monolayer hexagonal structure in which each C$_{60}$ cage binds to six neighbors, namely, a few-layer quasi-tetragonal phase (qTP) and a monolayer quasi-hexagonal phase (qHP) respectively. Great efforts have been devoted to stabilizing the linking bonds between neighboring cluster cages by introducing magnesium atoms to form a quasi-2D fullerene network with strong intralayer covalent bonds\,\cite{Hou2022} because Mg atoms tend to promote covalent bonds\,\cite{Pontiroli2013,Tanaka2018}. To aid exfoliation, the Mg ions that hold the C$_{60}$ cages together can be then replaced by large organic ions, which can be removed afterwards by hydrogen peroxide, leading to pure, charge neutral fullerene networks in 2D\,\cite{Hou2022,Gottfried2022}. Unfortunately, only qHP C$_{60}$ has been obtained as monolayers, while all the qTP C$_{60}$ flakes are few-layer\,\cite{Hou2022}. These results raise doubts regarding the stability of monolayer fullerene networks.

Ever since the discovery of C$_{60}$\,\cite{Kroto1985}, the formation mechanism and stability of the fullerene molecules are far from completely understood\,\cite{Kroto1987,Kroto1990,Goroff1996,Bernal2019}. When forming structural units of C$_{60}$ clusters in a 2D plane, it is unclear whether ordered structures of monolayer polymeric C$_{60}$ are stable under ambient conditions such as strain and temperature. Recent first principles calculations have investigated various structural phases of monolayer C$_{60}$ with different bonding characters\,\cite{Tromer2022,Yu2022,Yuan2022,Peng2022c,Ying2023,Dong2022}. The mechanical stability of several phases has been confirmed\,\cite{Yu2022,Tromer2022,Ying2023}. More recently, the thermal stability of monolayer C$_{60}$ has been addressed using molecular dynamics simulations, showing that both qTP and qHP C$_{60}$ monolayers can remain stable at temperatures near 800 K\,\cite{Ribeiro2022}, which is partially consistent with the experimental result that monolayer qHP C$_{60}$ does not decompose at 600 K\,\cite{Hou2022}. However, previous analysis based on mechanical and thermal stability cannot explain why the qTP monolayers have not yet been exfoliated experimentally. Furthermore, the dynamic stability of monolayer fullerene networks with respect to lattice vibrations, which indicates whether the crystal structure is in a local minimum of the potential energy surface\,\cite{Peng2017a,Malyi2019,Luo2022,Pallikara2022}, is still unexplored. Additionally, the thermodynamic stability of different phases, which energetically classifies the stability (especially at finite temperatures)\,\cite{Pavone1998,Pavone2001,Masago2006,Setten2007,Stoffel2010,Zhang2011a,Deringer2014,Deringer2014a,White2015,Nyman2016,Skelton2017,Pallikara2021,Bartel2022}, remains unknown. Because of such knowledge gaps, several questions need to be answered to understand the phase stability of monolayer fullerene networks: ($i$) Are qTP and qHP C$_{60}$, as pure carbon monolayers without extra Mg or organic ions to bind the C$_{60}$ cages together, dynamically stable? ($ii$) What is their relative stability in a thermodynamic aspect? ($iii$) Can the calculated mechanical strength support their phase stability?

\begin{figure*}
\begin{center}
\includegraphics[width=15cm]{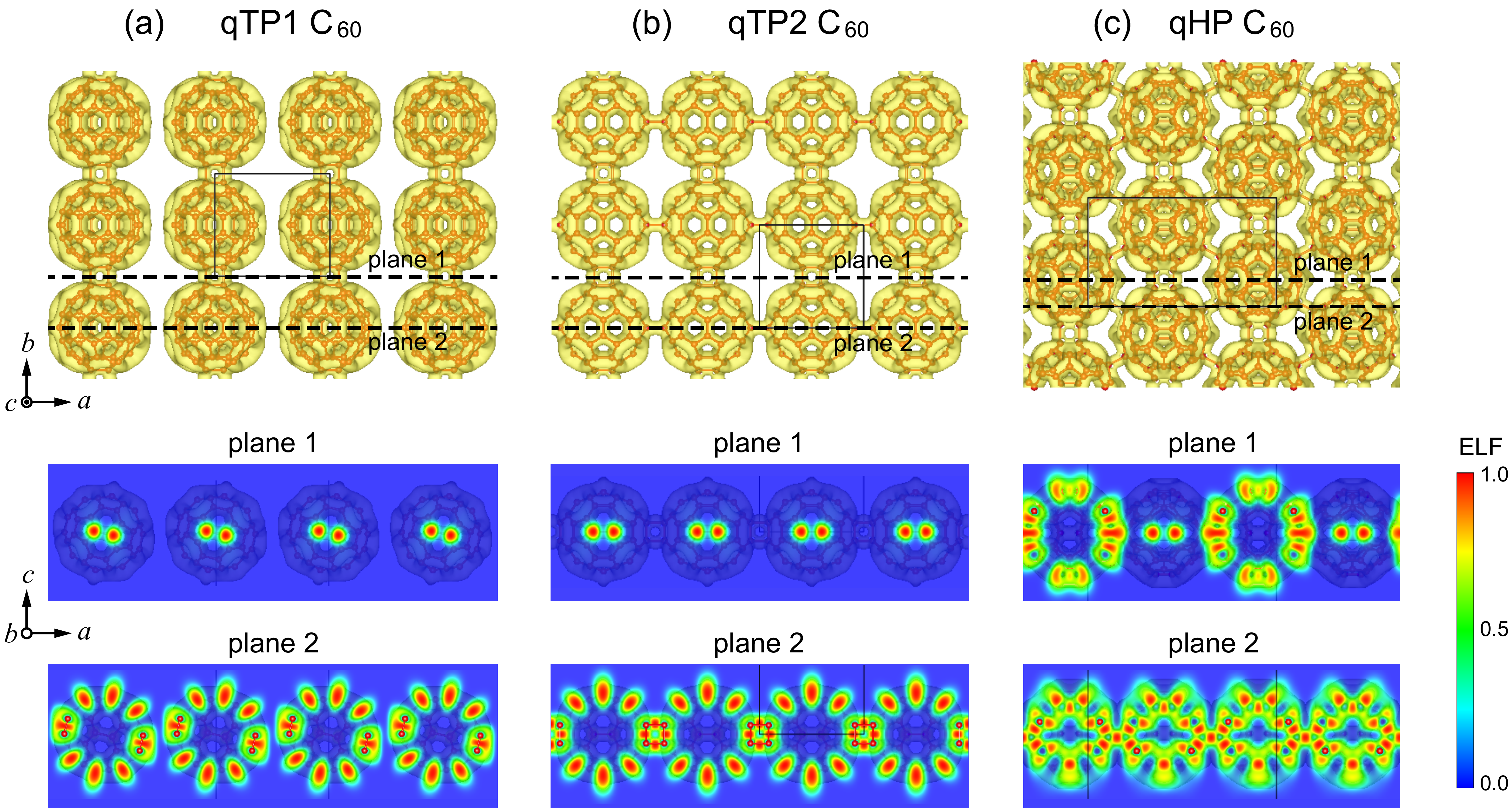}
\caption{Bond structures for (a) qTP1, (b) qTP2 and (c) qHP C$_{60}$. The default isosurface level in {\normalfont \scshape vesta}\,\cite{VESTA} is used. Maps of the ELF on the (010) plane are also present.}
\label{elf}
\end{center}
\end{figure*} 

In this work, I investigate the mechanical, dynamic and thermodynamic stability of monolayer qTP and qHP C$_{60}$ by using first principles calculations. Structural relaxation obtains two crystal structures of the quasi-tetragonal phase (denoted as qTP1 and qTP2 respectively). I show that the qTP1 monolayer, albeit being thermodynamically stable at all temperatures above 380 K, possesses low dynamic and mechanical stability due to its weak bonding perpendicular to the straight chains of C$_{60}$ buckyballs. On the other hand, although qTP2 fullerene might be the ground-state structure with the lowest Gibbs free energy at 0 K and exhibits good dynamic and mechanical stability, it is only thermodynamically stable with respect to qTP1 C$_{60}$ at low temperatures. Instead, monolayer qHP C$_{60}$ should be experimentally accessible owing to their dynamic and mechanical stability, in spite of its lowest thermodynamic stability among all three phases. In addition, qHP C$_{60}$ has the highest strength under various strains (hydrostatic, uniaxial and shear) because of the closely packed crystal structures. 

First principles calculations are performed using the Vienna \textit{ab-initio} simulation package ({\sc vasp})\,\cite{Kresse1996,Kresse1996a}. The projector augmented wave (PAW) potential is used with C $2s^22p^2$ valence states\,\cite{Bloechl1994,Kresse1999}, under the generalized gradient approximation (GGA) with the Perdew-Burke-Ernzerhof parameterization revised for solids (PBEsol) as the exchange-correlation functional\,\cite{Perdew2008}. The crystal structures are optimized by fully relaxing the lattice constants and internal atomic coordination (for computational details, see the Supporting Information). Geometry optimization starting from the quasi-tetragonal phase consisting of only carbon atoms leads to a quasi-1D qTP monolayer (qTP1), as shown in Fig.\,\ref{elf}(a). On the other hand, a two-step structural relaxation, starting from monolayer qTP Mg$_2$C$_{60}$ and then removing the Mg ions before the second relaxation, obtains a tightly bound qTP monolayer (qTP2), as shown in Fig.\,\ref{elf}(b). The two-step structure relaxation mimics the experimental procedure to remove the charged ions introduced during synthesis\,\cite{Hou2022,Gottfried2022}. The computed lattice constants for all three phases are listed in Table\,\ref{lattice}, which are in good agreement with previous results\,\cite{Yu2022,Tromer2022,Dong2022}, therefore confirming the reliability of the present calculations.

\begin{table}
	\begin{center}
	\caption{Calculated static lattice constants (in \AA) and cohesive energy $E_{\mathrm{c}}$ (in eV/atom) 
	of qTP1, qTP2 and qHP C$_{60}$ monolayers, 1D qTP C$_{60}$ chain and 0D C$_{60}$ molecule. The cohesive energy is defined as $E_{\mathrm{c}} = E_{\mathrm{ tot}} / N - E_{\mathrm{ isolated}}$, where $E_{\mathrm{ tot}}$ is the total energy of the crystal, $N$ is the number of atoms in the unit cell and $E_{\mathrm{ isolated}}$ is the total energy of an isolated carbon atom. The room-temperature lattice constants calculated under the quasi-harmonic approximation are also listed in parentheses for comparison.
	}
	\label{lattice}
		\begin{tabular}{lccc}	
\hline\hline		
phase & $a$ & $b$ & $E_{\mathrm{c}}$  \\
\hline
2D qTP1 &   10.491 	&   9.063   &  --9.2582 \\  
 &   (10.522)	&   (9.090)  &   \\  
2D qTP2 &    9.097 	&   9.001   &  --9.2587 \\  
  &    (9.132)	&   (9.031)  &   \\  
2D qHP  &   15.848 	&   9.131   &  --9.2465 \\  
  &   (15.896)	&   (9.162)  &   \\  
 1D  &    9.062  &   -      &  --9.2579 \\  
    &    (9.098)	&   -      &   \\  
 0D  &    -  	&   -      &  --9.2564 \\  
\hline\hline
	\end{tabular}
	\end{center}
\end{table}

\begin{table*}
	\begin{center}
	\caption{Elastic properties for qTP1, qTP2 and qHP C$_{60}$, with the elastic constants $C_{ij}$, shear modulus $G^{\mathrm{2D}}$, layer modulus $\gamma$, Young's modulus $Y^{\mathrm{2D}}$ in N/m, and Poisson's ratio $\nu$ dimensionless. The elastic constants $C_{11}$, $C_{22}$ and $C_{66}$ calculated from phonon speed of sound are also listed in parentheses for comparison.}
	\label{elastic}
		\begin{tabular}{lccccccccc}	
\hline\hline		
phase & $C_{11}$ & $C_{22}$ & $C_{12}$ & $C_{66}=G^{\mathrm{2D}}$ & $\gamma$ & $Y^{\mathrm{2D}}_a$ & $Y^{\mathrm{2D}}_b$ & $\nu_a$ & $\nu_b$ \\
\hline
qTP1 & 5.4 & 123.7 & --1.2 & --0.2 & 31.7 & 5.4 & 123.5 & --0.010 & --0.225 \\
 & (2.5) & (121.3) & - &  \\
qTP2 & 149.9 & 148.7 & 22.9 & 53.4 & 86.1 & 146.4 & 145.2 & 0.154 & 0.153 \\
 & (150.5) & (141.2) & - & (54.5) \\
qHP  & 150.8 & 186.8 & 22.5 & 60.6 & 95.6 & 148.1 & 183.4 & 0.120 & 0.149 \\
 & (142.4) & (172.7) & - & (61.7) \\
\hline\hline
	\end{tabular}
	\end{center}
\end{table*}

The bond structures at equilibrium are examined in Fig.\,\ref{elf}. The relaxed structure for qTP1 fullerene can be regarded as one-dimensional chains of C$_{60}$ cages along the $b$ direction that are linked by the nearly in-plane [2+2] cycloaddition bonds. In comparison, qTP2 fullerene is a two-dimensional network of C$_{60}$ cages connected by the out-of-plane vertical [2+2] cycloaddition bonds along the $a$ direction and the in-plane [2+2] cycloaddition bonds along the $b$ direction. The major difference between qTP1 and qTP2 is the absence of the vertical [2+2] cycloaddition bonds along $a$ in the former. Regarding the qHP monolayer, the C$_{60}$ cages form a hexagonal network through the similar planar [2+2] cycloaddition bonds along the $b$ direction and C$-$C single bonds along the other two directions diagonal to the rectangular unit cell.

Figure\,\ref{elf} also shows the maps of the electron localization function (ELF) on the (010) plane. A high value of ELF indicates strong electron localization\,\cite{Becke1990,Savin1992,Gatti2005,Chen2013}. As shown in Fig.\,\ref{elf}(a), the covalent [2+2] bonds along $b$ in qTP1 fullerene lead to high electron localization there (plane 1), whereas no bonds are formed between neighboring C$_{60}$ cages along $a$ (plane 2). In contrast, the vertical [2+2] bonds along $a$ in qTP2 fullerene result in high electron localization between neighboring C$_{60}$ cages, as demonstrated in plane 2 of Fig.\,\ref{elf}(b). For qHP C$_{60}$, the hexagonal network has higher electron localization in both directions, as one can see from Fig.\,\ref{elf}(c). As a result, one can expect that the hexagonal networks should stabilize and strengthen the structure of qHP C$_{60}$, making it slightly more stable than qTP2 C$_{60}$ while much more stable than qTP1 C$_{60}$. However, as shown below, although the mechanical and dynamic stabilities are consistent with the ELF picture, high electron localization in qHP C$_{60}$ does not guarantee their thermodynamic stability.

To confirm the mechanical stability, the elastic constants are calculated by finite differences through finite distortions of the lattice\,\cite{LePage2002,Wu2005}. There are different ways to define the 2D elastic constants from the computed 3D coefficients\,\cite{Blonsky2015,Peng2017a,Pashartis2022}. Here the 2D coefficients $C_{ij}^{\mathrm{2D}}$ are renormalized by the $c$ lattice constant (the spacing between 2D layers)\,\cite{Blonsky2015,Peng2017a}, i.e. $C_{ij}^{\mathrm{2D}} = c \times C_{ij}^{\mathrm{3D}}$. The obtained 2D elastic constants (including ionic relaxations) are listed in Table\,\ref{elastic} using the Voigt notation: $1-xx$, $2-yy$, $6-xy$, and the present results agree well with previous calculations\,\cite{Yu2022,Tromer2022}. According to Born-Huang's lattice dynamical theory\,\cite{Born1954,Wu2007}, in monoclinic crystals (qTP1 and qHP with space group $P2/m$ and $Pc$ respectively), the mechanical stability criteria is given by
\begin{equation}
\label{Born}
    C_{11} > 0,\ C_{22} > 0,\ C_{66} > 0,\ C_{11} + C_{22} + 2C_{12} > 0.
\end{equation}
In orthorhombic crystals (qTP2 C$_{60}$ with space group $Pmmm$), the Born stability criteria has an extra requirement in addition to Eq.\,(\ref{Born})
\begin{equation}
    C_{11} + C_{22} - 2C_{12} > 0.
\end{equation}
The elastic constants of qTP2 and qHP C$_{60}$ satisfy their corresponding criteria, indicating that they are mechanically stable. Interestingly, the shear strength $G^{\mathrm{2D}}$ of qTP1 C$_{60}$ is negative, demonstrating its low shear resistance. The 1D chains in qTP1 C$_{60}$ are prone to bending under shear deformation, which may lead to a sliding of C$_{60}$ chains and even lattice instability. In addition, $C_{11}$ in qTP1 fullerene is more than one order of magnitude lower than $C_{22}$. Such weak stiffness is correlated to weak interchain bonding effect along $a$ as discussed above and the weak dynamic stability as will be demonstrated below. In contrast, $C_{11}$ and $C_{22}$ are nearly the same in qTP2 fullerene because of the similar [2+2] cycloaddition bonds along both $a$ and $b$. For qHP fullerene, the elastic constants $C_{11}$, $C_{22}$ and $C_{66}$ are the highest among the three phases, in consistent with the high electron localization in the hexagonal networks that strengthens the crystal structure.

The strength of monolayer fullerene networks is obtained from the computed elastic constants, as summarized in Table\,\ref{elastic}. The layer modulus $\gamma$ is the 2D equivalent to the bulk modulus, which measures the resistance to hydrostatic stretching in 2D materials\,\cite{Andrew2012}. The layer moduli show an increasing trend from qTP1 to qTP2 and to qHP C$_{60}$. The $\gamma$ for qTP2 C$_{60}$ is more than twice that of qTP1 C$_{60}$, while slightly lower than that of qHP C$_{60}$, which concurs with the bonding structures of three fullerene networks. 
In general, qTP2 and qHP C$_{60}$ have comparable moduli and therefore similar hardness properties, whereas qTP1 C$_{60}$ has less resilience to both shear and hydrostatic strains.

\begin{figure*}
\begin{center}
\includegraphics[width=17.4cm]{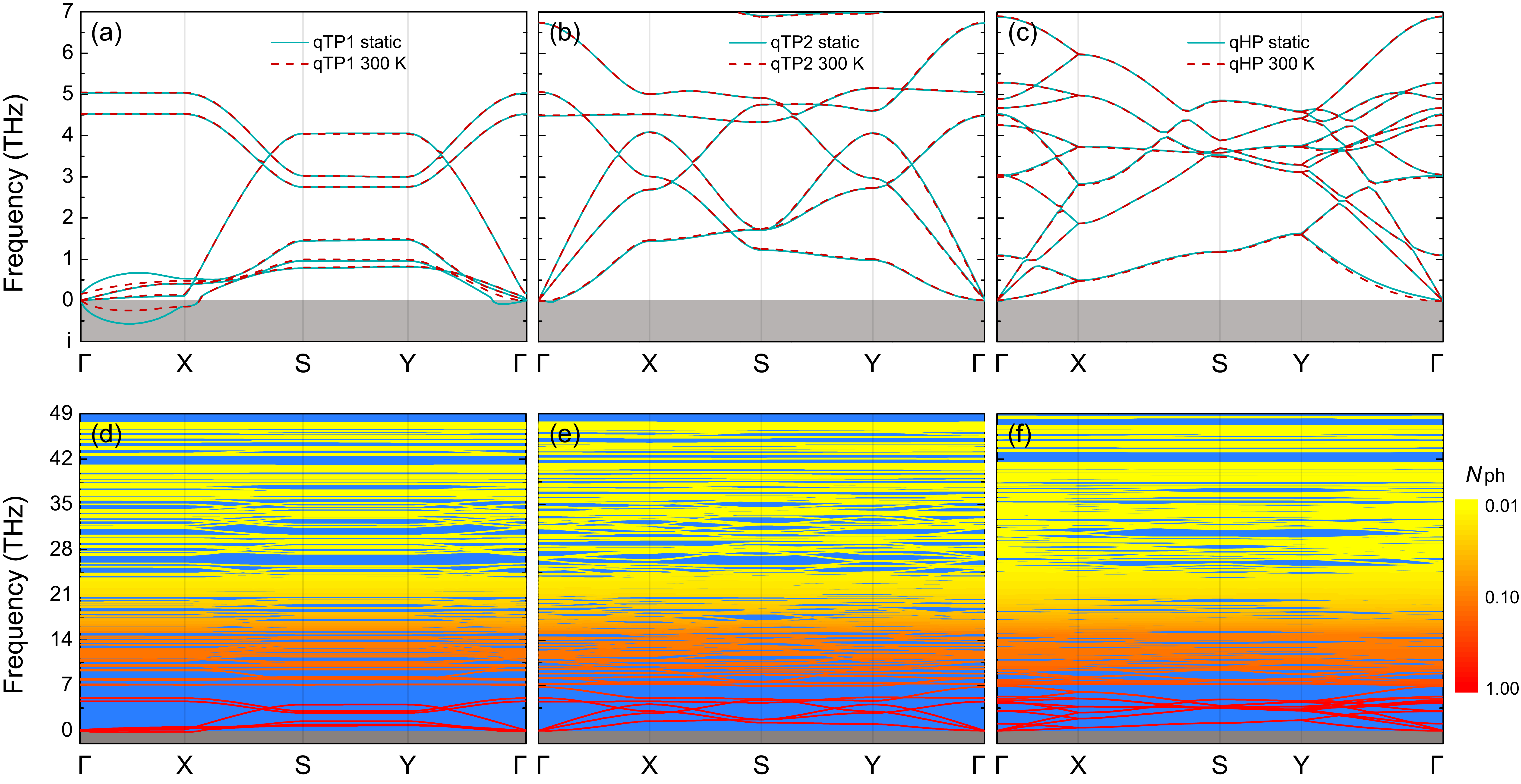}
\caption{Low-frequency phonons of (a) qTP1, (b) qTP2 and (c) qHP C$_{60}$ using the static and room-temperature lattice constants. Entire phonon spectra for (d) qTP1, (e) qTP2 and (f) qHP C$_{60}$ using the room-temperature lattice constants, with the phonon occupation number $N_{\mathrm{ph}}$ determined from the Bose-Einstein distribution function at 300 K.}
\label{phonon}
\end{center}
\end{figure*} 

The anisotropy of strength is also investigated by calculating the Young's modulus and Poisson's ratio. In qTP1 fullerene, the Young's modulus $Y^{\mathrm{2D}}$ along $a$ is more than 22 times lower than that along $b$, indicating that qTP1 C$_{60}$ is much less structurally rigid to elongations along $a$. In qTP2 C$_{60}$, the Young's moduli along $a$ and $b$ are nearly the same due to similar [2+2] cycloaddition bonds along both directions, showing that they have the same resilience to linear strain. Regarding the qHP monolayers, the $Y^{\mathrm{2D}}_a$ has a value 80\% the $Y^{\mathrm{2D}}_b$, indicating slightly weaker stiffness of the C$-$C single bonds in the presence of strain along $a$. 
The Poisson's ratio $\nu$ for qTP1 C$_{60}$ is negative, i.e. the qTP1 fullerene monolayers expand laterally when stretched, and the $ | \nu_a | $ is significantly lower than $ | \nu_b | $ because of much less bond stretching under uniaxial strain. Monolayer qTP2 C$_{60}$ has a nearly isotropic $\nu$ of $0.153-0.154$, while the $\nu_a$ in qHP fullerene is slightly lower than $\nu_b$. These results indicate that qTP1 C$_{60}$ is unable to withstand greater strains along $a$ than those along $b$, which is the origin of its overall low strength.

To evaluate the dynamic stability of monolayer fullerene networks, lattice dynamical properties are calculated within the harmonic approximation based on density functional perturbation theory\,\cite{DFPT,Gonze1995,Gonze1995a}. The phonon spectra of all three phases are gathered in Fig.\,\ref{phonon}. 
As shown in Fig.\,\ref{phonon}(a), the phonon dispersion of qTP1 C$_{60}$ using the static lattice constants exhibits small imaginary frequency ($<$ 0.6i THz) along the entire $\Gamma$-X high-symmetry line. An imaginary frequency indicates a decrease in potential energy when the atoms are displaced away from their equilibrium positions, corresponding to a non-restorative force\,\cite{Pallikara2022}. Therefore, the imaginary frequency along $\Gamma$-X implies that monolayer qTP1 C$_{60}$ can be split into individual 1D chains in the presence of interchain (out-of-plane) vibrations, demonstrating its weak dynamic stability along the $a$ direction. There is a fourth mode at $\Gamma$ with nearly zero frequency in qTP1 C$_{60}$, which, sometimes known as the torsional acoustic mode, is a strong indication of the (quasi-)1D nature\,\cite{Peng2018,Lange2022}. The thermal expansion is included by computing the Gibbs free energy under the quasi-harmonic approximation\,\cite{Huang2015a,Huang2016,Peng2019}, and the room-temperature lattice constants are listed in Table\,\ref{lattice}. At 300 K, the imaginary mode in qTP1 C$_{60}$ remains along $\Gamma$-X, though the imaginary frequency becomes smaller ($<$ 0.2i THz). 
In contrast, qTP2 and qHP fullerene are dynamically more stable as there is no imaginary mode in Fig.\,\ref{phonon}(b) and (c) using both the static and room-temperature lattice constants, indicating these structures are a local minimum on the potential energy surface and the atoms vibrate harmonically around their equilibrium positions.

From the phonon speed of sound, the elastic constants $C_{11}$, $C_{22}$ and $C_{66}$ can be calculated\,\cite{Feng2012} (for details on the phonon group velocity, see the Supporting Information). As shown in Table\,\ref{elastic}, the calculated elastic constants are in reasonably good agreement with those computed from the finite difference method\,\cite{LePage2002,Wu2005}. Moreover, qHP fullerene has the highest speed of sound along $b$ and the highest phonon frequency with four phonon branches higher than 48.4 THz throughout the entire Brillouin zone, whereas the highest phonon frequencies in qTP1 and qTP2 fullerene are lower than 47.7 THz, which is in line with the high mechanical strength in qHP C$_{60}$.

To clarify the thermodynamic stability of C$_{60}$ monolayers, the cohesive energy is calculated, as listed in Table\,\ref{lattice}. The resulting $E_{\mathrm{c}}$ of qTP2 fullerene is 0.5 meV/atom (30 meV per formula unit) lower than qTP1 and 12.2 meV/atom (732 meV per formula unit) lower than qHP, suggesting its thermodynamic stability. However, because the energy difference between the qTP1 and qTP2 monolayers is quite small, phonons can play an important role in determining the thermodynamic stability at both 0 K and finite temperatures\,\cite{Pavone1998,Pavone2001,Masago2006,Setten2007}. The contribution of phonons can be examined by calculating the Gibbs free energy $F$\,\cite{Dove1993,Togo2008,Togo2015}
	\begin{equation}
	\label{helmholtz}
			F= \mathop{\min}_{a,b} \bigg[
			E_{\mathrm{ tot}}
			+\frac{1}{2}\sum_{\lambda}\hbar \omega_{\lambda}
			+k_BT\sum_{\lambda}
			\ln{\big( 1- \mathrm{e} ^{-\hbar \omega_{\lambda}/k_BT} \big)
			}
			\bigg]
			,
	\end{equation}
where $\mathop{\min}_{a,b} [\ ]$ finds unique minimum value in the brackets by changing the lattice constants $a$ and $b$ to include thermal expansion, $E_{\mathrm{ tot}}$ is the total energy of the crystal, $\hbar$ is the reduced Planck constant, $\omega_{\lambda}$ is the phonon frequency at mode $\lambda$, $k_B$ is the Boltzmann constant, and $T$ is the temperature. The second term in Eq.\,(\ref{helmholtz}) is temperature-independent, corresponding to the zero point energy (ZPE) of phonons; and the last term refers to the thermally excited population of phonons, as demonstrated by the Bose-Einstein distribution $N_{\mathrm{ph}}$ at 300 K in Fig.\,\ref{phonon}(d)-(f). 

\begin{figure}
\begin{center}
\includegraphics[width=7cm]{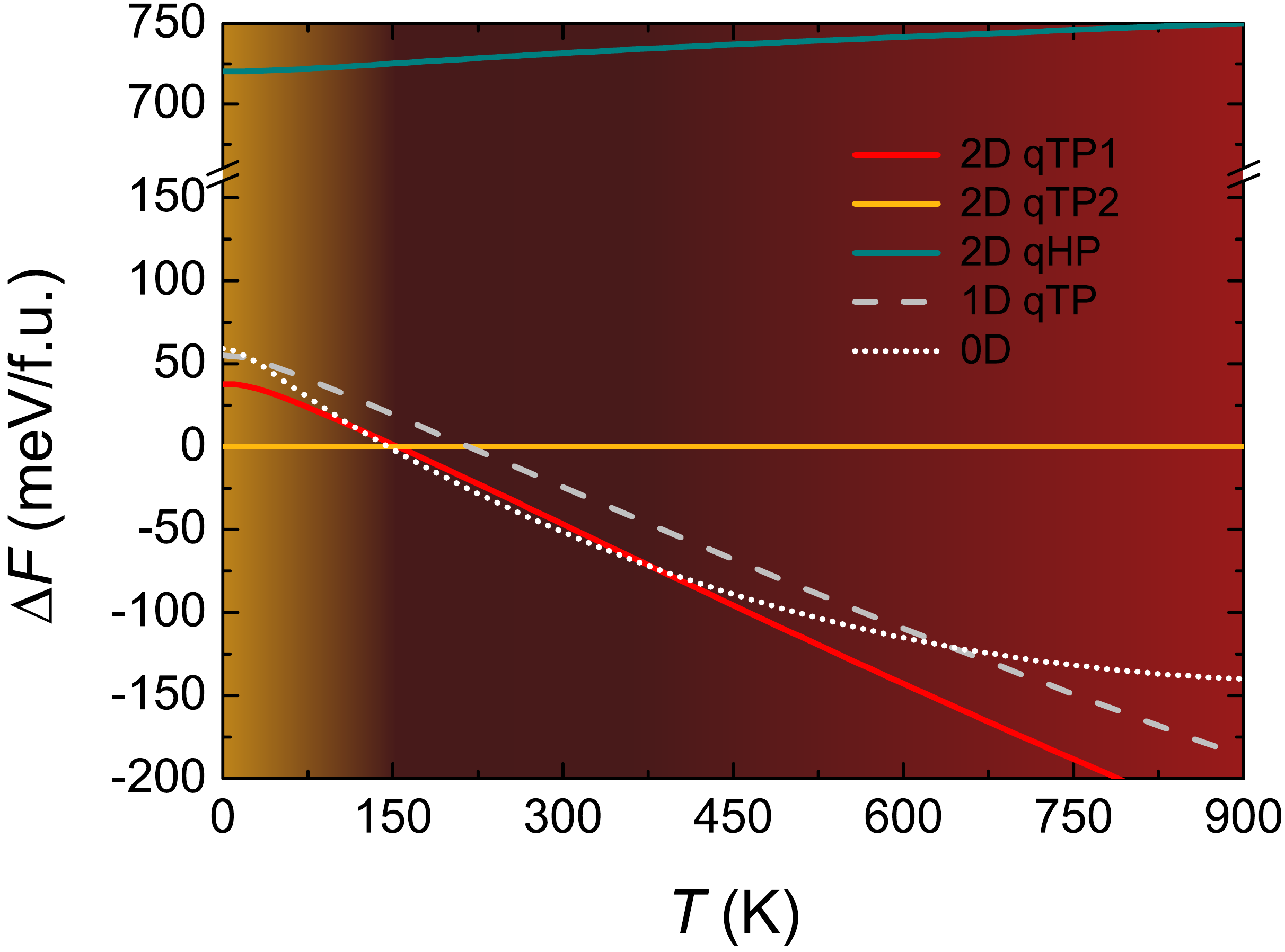}
\caption{
Relative thermodynamic stability of monolayer fullerene networks, one-dimensional fullerene chain and zero-dimensional fullerene molecule, with the Gibbs free energy $F$ of monolayer qTP2 C$_{60}$ set to zero to compare the relative stability.
}
\label{free}
\end{center}
\end{figure} 

To quantify the relative thermodynamic stability at finite temperatures, the difference in Gibbs free energy $\Delta F$ with respect to the free energy of monolayer qTP2 C$_{60}$ is plotted as a function of temperature $T$ for all three phases, as illustrated in Fig.\,\ref{free}. With increasing temperature, the free energy of qTP1 C$_{60}$ drops faster than that of qTP2 C$_{60}$ owing to their smaller vibrational frequencies. According to Eq.\,(\ref{helmholtz}), smaller vibrational frequencies give rise to lower free energy and higher entropy (for details on the phonon density of states and entropy, see the Supporting Information), which is similar to the case in $\alpha$- and $\beta$-tin\,\cite{Pavone1998,Pavone2001}. At 150 K, monolayer qTP1 C$_{60}$ becomes thermodynamically more stable than the other two phases. At 300 K, the free energy of qTP1 C$_{60}$ lies 47 meV per formula unit (meV/f.u.) below that of qTP2 C$_{60}$. At higher temperatures, the difference becomes even larger, further stabilizing the qTP1 structure in a thermodynamic perspective.

To further explore the thermodynamic stability of C$_{60}$ in different dimensions, the monolayer qTP1 network is further isolated into 1D qTP C$_{60}$ chain and 0D C$_{60}$ molecule (for details about 1D and 3D C$_{60}$, see the Supporting Information). The cohesive energy $E_{\mathrm{c}}$ of 1D qTP C$_{60}$ and 0D C$_{60}$ is listed in Table\,\ref{lattice}. Interestingly, the $E_{\mathrm{c}}$ of 1D qTP C$_{60}$ chain is higher than monolayer qTP1 and qTP2 C$_{60}$ network by merely 0.3 and 0.8 meV/atom respectively (18 and 48 meV/f.u.), whereas 11.4 meV/atom (684 meV/f.u.) lower than the $E_{\mathrm{c}}$ of 2D qHP C$_{60}$. The difference in $E_{\mathrm{c}}$ between 1D qTP C$_{60}$ chain and monolayer qTP1 C$_{60}$ network is even lower than the thermal fluctuation energy $k_{\mathrm{B}}T$ at room temperature (26 meV), implying that 2D qTP1 C$_{60}$ can be transformed into 1D chains in the presence of thermal fluctuations. Taking the finite temperature effects into account, the Gibbs free energy of 1D qTP C$_{60}$ chain and the Helmholtz free energy of 0D C$_{60}$ molecule are shown in Fig.\,\ref{free} as a function of temperature. The Gibbs free energy of 1D qTP C$_{60}$ chain is higher than that of 2D qTP1 C$_{60}$ in the entire temperature range ($0-900$ K), and their free energy difference is 22 meV/f.u. at 300 K. On the other hand, the free energy of 1D qTP C$_{60}$ chain becomes lower than that of 2D qTP2 C$_{60}$ at temperatures above 220 K. Most interestingly, the free energy of 0D C$_{60}$ molecule drops faster than all the other phases below room temperature, and becomes lower than 2D qTP1 and qTP2 C$_{60}$ at 120 and 150 K respectively. However, the free energy of 2D qTP1 C$_{60}$ decreases the fastest above room temperature, and consequently 2D qTP1 C$_{60}$ is energetically more favored than all the other phases at temperatures above 380 K. As a result, monolayer qTP2 C$_{60}$ is thermodynamically most stable at temperatures below 150 K, 0D C$_{60}$ molecule has the lowest energy for temperatures between 150 and 380 K, and 2D qTP1 C$_{60}$ is thermodynamically favored above 380 K. 





Looking back at the calculated mechanical properties and stabilities, they seem in line with the experimental findings. It has been reported that fullerene monolayers can only be isolated experimentally for the honeycomb structure qHP, whereas the obtained rectangular structure qTP is few-layered\,\cite{Hou2022}. Although qTP2 C$_{60}$ is thermodynamically favored over qTP1 C$_{60}$ at low temperatures, clearly qTP1 C$_{60}$ is thermodynamically more stable than the other two phases at all temperatures above 150 K. 
However, thermodynamic stability of qTP1 C$_{60}$ does not guarantee high dynamic stability in the presence of interchain (out-of-plane) vibrations perpendicular to the quasi-1D chains. 
In addition, the low moduli and strength of qTP1 C$_{60}$ originated from the chain crystal structures, in addition to its low shear resistance, indicate that qTP1 C$_{60}$ cannot be intrinsically resilient under deformation. Moreover, monolayer qTP1 C$_{60}$ is thermodynamically less stable than 0D C$_{60}$ molecule for temperatures between 120 and 380 K. These results indicate the plausibility that monolayer qTP1 fullerene network can be further split into individual 1D chains or 0D molecules by thermal fluctuations, interchain acoustic vibrations, or external strains. 
In contrast, qHP C$_{60}$ is both dynamically and mechanically more stable with respect to qTP1 C$_{60}$. Therefore, monolayer polymeric C$_{60}$ has so far only been exfoliated from the quasi-hexagonal bulk single crystals. These results indicate that a systematic analysis of mechanical, dynamic and thermodynamic stabilities is necessary to rationalize the experimental data.





In conclusion, I carry out first principles calculations to evaluate the mechanical, dynamic and thermodynamic stabilities of qTP1, qTP2 and qHP C$_{60}$ monolayers, which have been so far believed to be stable. Electron localization analysis reveals that the low mechanical and dynamic stabilities in qTP1 fullerene are associated with the lack of C$-$C bonds connecting the adjacent C$_{60}$ chains, which also limits its achievable strength. Monolayer qTP2 C$_{60}$ is thermodynamically more stable at temperatures below 150 K, while thermally populated phonons hinder its thermodynamic stability with increasing temperature. The relatively high moduli of qHP fullerene indicate that it has a high strength 
because of the closely packed hexagonal fullerene network linked through both [2+2] cycloaddition bonds and C$-$C single bonds. This, in combination with the phonon stability, endows monolayer qHP C$_{60}$ with high stability and strength.

\section*{Acknowledgements}

I thank Prof. Bartomeu Monserrat and Dr. Ivona Bravi\'{c} at the University of Cambridge for helpful discussions. I acknowledge support from the Winton Programme for the Physics of Sustainability, and from Magdalene College Cambridge for a Nevile Research Fellowship. The calculations were performed using resources provided by the Cambridge Tier-2 system, operated by the University of Cambridge Research Computing Service (www.hpc.cam.ac.uk) and funded by EPSRC Tier-2 capital grant EP/P020259/1, as well as with computational support from the U.K. Materials and Molecular Modelling Hub, which is partially funded by EPSRC (EP/P020194), for which access is obtained via the UKCP consortium and funded by EPSRC grant ref. EP/P022561/1.



\providecommand*\mcitethebibliography{\thebibliography}
\csname @ifundefined\endcsname{endmcitethebibliography}
  {\let\endmcitethebibliography\endthebibliography}{}

\end{document}